\documentclass[printer]{aa} % for two columns
\bibpunct{(}{)}{;}{a}{}{,} % to follow the A&A style
\usepackage{graphicx}
\usepackage{txfonts}
\begin{document}

   \title{Detection of thermal radio emission from a single coronal giant}

   \author{E. O'Gorman
          \inst{1}\fnmsep\thanks{ogorman@cp.dias.ie},
          G. M. Harper
          \inst{2},
          \and
          W. Vlemmings
          \inst{3}
			}
   \institute{Dublin Institute for Advanced Studies, 31 Fitzwilliam Place, Dublin 2, Ireland
          \and
             Center for Astrophysics and Space Astronomy, University of Colorado, 389 UCB, Boulder, CO 80309, USA
           \and     
             Department of Earth and Space Sciences, Chalmers University of Technology, Onsala Space Observatory, 439 92 Onsala, Sweden
             }
            
  \abstract{We report the detection of thermal continuum radio emission from the K0\,III coronal giant Pollux ($\beta$ Gem) with the Karl G. Jansky Very Large Array (VLA). The star was detected at 21 and 9\,GHz with flux density values of $150\pm21$ and $43\pm8\,\mu$Jy, respectively. We also place a $3\sigma _{\mathrm{rms}}$ upper limit of $23\,\mu$Jy for the flux density at 3\,GHz.  We find the stellar disk-averaged brightness temperatures to be approximately 9500, 15000, and $<71000\,$K, at 21, 9, and 3\,GHz, respectively, which are consistent with the values of the quiet Sun. The emission is most likely dominated by optically thick thermal emission from an upper chromosphere at 21 and 9\,GHz. We discuss other possible additional sources of emission at all frequencies and show that there may also be a small contribution from gyroresonance emission above active regions, coronal free-free emission and free-free emission from an optically thin stellar wind, particularly at the lower frequencies. We constrain the maximum mass-loss rate from Pollux to be less than $3.7\times 10^{-11}\,M _{\odot}$\,yr$^{-1}$ (assuming a wind terminal velocity of 215\,km\,s$^{-1}$), which is about an order of magnitude smaller than previous constraints for coronal giants and is in agreement with existing predictions for the mass-loss rate of Pollux. These are the first detections of thermal radio emission from a single (i.e., non-binary) coronal giant and demonstrate that low activity coronal giants like Pollux have atmospheres at radio frequencies akin to the quiet Sun.}
   \titlerunning{Detection of thermal radio emission from a single coronal giant}
   \authorrunning{E. O'Gorman et al.}
   \maketitle

\section{Introduction}
Unlike low mass stars, moderate-mass ($M_{\star} \sim 3\,M_{\odot}$) late B and early A spectral type stars lose little angular momentum during their main-sequence (MS) phase \citep{wolff_1997}. Once shell hydrogen burning commences, these stars move away from the MS and begin to form surface convection zones while evolving rapidly through the F and G spectral types. As the surface convection zone deepens, coronal activity is activated and the star ascends the red giant branch (RGB). Subsequently it returns close to the base of the RGB where it becomes a `clump' giant burning helium in its core \citep{ayres_1998}. This coronal activity will inevitably subside as the star spins down. Therefore, late B and early A spectral type stars which do not have coronae on the MS, will develop coronae during a period of their post MS evolution. This is in contrast to lower mass stars like the Sun, which display strong coronal emission on the MS due to intense rotation-induced dynamo-driven magnetic activity \citep{parker_1970}, which subsides as they evolve to red giants. 

The atmospheres of coronal giants have been extensively studied at both X-ray \citep{haisch_1991, maggio_1990, huensch_1996} and ultra-violet (UV) wavelengths \citep{ayres_2003, dupree_2005}. X-ray studies have revealed that stellar coronae are common in the giant branch for spectral types earlier than K1\,III but become rare redward of K1\,III, a region which has thus been dubbed the `coronal graveyard' \citep{ayres_1991}. The UV emission line studies agree with these findings but also suggest pervasive transition region material at $10^5$\,K among many of the non-coronal giants \citep{ayres_1997}. 

Detecting radio emission from coronal giants provides an alternative method to study their atmospheres and allows a comparison with the atmospheres of coronal MS stars. Active coronal MS stars can have intense radio luminosities many orders of magnitude greater than that of the disk-averaged quiet (i.e., active region free) solar value. These high values have been associated with gyrosynchrotron emission from a continuously generated population of non-thermal electrons in their coronae \citep{linsky_1983}. Less active coronal MS stars, like the Sun, emit only thermal emission at radio frequencies in their quiescent (i.e., non-flaring) states which is entirely due to free-free processes. Subsequently, their radio luminosities are much lower than those from more active coronal MS stars and only recently has this thermal radio emission been detected from MS stars other than the Sun \citep{villadsen_2014}. Unlike coronal MS stars, little is known about the radio atmospheres of coronal giants. \cite{drake_1986} observed a number of nearby coronal giants at 5\,GHz (i.e., 6\,cm) with the Very Large Array. They failed to detect any single coronal giants but did detect the binary Capella ($\alpha$ Aur; G5\,III + F9\,III), which is a long period  RS CVn type system. The radio emission from this system was spatially unresolved and was interpreted to be thermal free-free emission from one or both of their chromospheres and coronae.

In this paper we report radio observations of the closest coronal giant, Pollux ($\beta$ Gem; K0\,III), whose basic properties are summarized in Table \ref{tab1}. According to \cite{auriere_2009}, Pollux is probably a descendent of an early-A spectral type MS star and now lies either at the base of the red giant branch or is a clump giant. Pollux is a weakly-active magnetic giant and a sub-Gauss value has been obtained for its surface averaged longitudinal magnetic field \citep{auriere_2009}. It has an X-ray luminosity that is 3 orders of magnitude lower than Capella's \citep{huensch_1996}, despite its similar size to both of the stars in the Capella binary. Extensive studies of Pollux's radial velocity variations have revealed it to have a planetary companion with a period of 560 days and a minimum mass of 2.9\,$M_\mathrm{Jup}$ \citep{reffert_2006, hatzes_2006}. Understanding the radio properties of the stellar host of an exoplanet not only provides insight on the host itself but also provides insight on the feasibility of detecing radio emission from the exoplanet itself.

\begin{table}
\caption{Basic stellar properties of Pollux ($\beta$ Gem)}
\label{tab1}
\centering
\begin{tabular}{l l c}
\hline\hline
\rule{-2.6pt}{2.5ex}Parameter & Value & Reference \\
\hline\hline
\rule{-2.6pt}{2.5ex} HD number\dotfill  										& 62509 & - \\
                     Spectral type\dotfill  									& K0\,III & 1 \\
					  Effective temperature $T_{\mathrm{eff}}$ \dotfill   		& $4904\pm 50\,$K & 2\\	
					  Distance $d$ \dotfill										& $10.36\pm 0.03\,$pc & 3\\
					  Angular diameter $\phi _{\star}$ \dotfill	    		& $7.95\pm 0.09\,$mas & 4\\	
					  Mass $M_{\star}$ \dotfill									& $1.91\pm 0.09\,M_{\odot}$ & 5\\									  
					  Radius $R_{\star}$ \dotfill								& $8.85\pm 0.10\,R_{\odot}$ & c \\
					  $[$Fe/H$]$	 \dotfill													& 0.08						& 1 \\
  					  Escape velocity $v_{\star}^{\mathrm{esc}}$ \dotfill 				& $287\,$km\,s$^{-1}$ & c \\
  					  Age 		 \dotfill										& $1.2 \pm 0.3 $\,Gyr	   & 5 \\
  					  Soft X-ray luminosity $L_\mathrm{x}$ \dotfill 				& $10^{27.13}$\,erg\,s$^{-1}$ & 7 \\   				  				  					 
\hline
\end{tabular}
      \vspace{-2mm}
     \tablefoot{(1) \cite{gray_2003}; (2) \cite{takeda_2008}; (3) \cite{van_leeuwen_2007}; (4) \cite{nordgren_2001}; (5) \cite{hatzes_2012}; (6) \cite{gray_2014}; (7) \cite{sanz_forcada_2011}. `c' indicates values calculated in this work.}
\end{table}

\section{Observations and data reduction}\label{sec2}
Pollux was observed with the Karl G. Jansky Very Large Array (VLA) in A configuration (Program code: 12B-108, PI: Laurent Chemin) during 2012 October at K band ($18.0 - 26.5\,$GHz) and again during 2013 January at both X band ($8.0 - 12.0$\,GHz) and S band ($2.0 - 4.0\,$GHz). The original goal of this project was to test whether the signature of Pollux's exoplanet could be detected in its orbital reflex motion with a future astrometric Very Long Baseline Array (VLBA) program. The low levels of radio flux density that we report in this paper would appear to rule out such a possibility with current facilities. A brief overview of these observations is provided in Table \ref{tab2}. The receivers recorded continuum emission with 8-bit samplers in 2\,GHz wide windows centered at 21.2 (K band), 9.0 (X band), and 3.0\,GHz (S band). For all frequency setups, 3C286 was used as the absolute flux density and bandpass calibrator, while J0741+3112 was used as the complex gain calibrator. Pollux was observed offset $\sim 8^{\prime \prime}$ due west of the phase reference center and so any potential interferometric artefacts at phase center could not be mistaken for the source. 

Flagging and calibration were performed with the Common Astronomy Software Applications (CASA) package \citep{mcmullin_2007} version 4.2.2 using the VLA calibration pipeline version 1.3.1 and some manual flagging. Images were created using the CLEAN algorithm \citep{hogbom_1974} in full Stokes, with Briggs weighting and a robust parameter of 0.5. At S band, we were required to image a number of bright nearby (i.e., a few arcminutes) serendipitous sources to suppress their sidelobe contamination throughout the image. In the final primary beam-corrected images, the rms noise per synthesized beam close to the source was 14$\,\mu$Jy beam$^{-1}$ at K band (21.2\,GHz), 6$\,\mu$Jy beam$^{-1}$ at X band (9.0\,GHz), and 7.5$\,\mu$Jy beam$^{-1}$ at S band (3.0\,GHz). The source flux density and position were determined by fitting a point source to the calibrated visibilities using the CASA task \textit{uvmodelfit}. To mitigate the possibility of decorrelation of the visibilities at 21.2\,GHz, which could result in flux scale errors, we limited the point source fit to baselines less than 500\,k$\lambda$ at this frequency. The flux density uncertainty for standard VLA observations is $\sim 3\%$ at S and X band, and $\sim 10\%$ at K band. The error bars throughout this paper do not include this uncertainty in the absolute flux calibration but reflect both the statistical root-mean-square (rms) error values, $\sigma _{\mathrm{s}}$, obtained from the final primary beam corrected images, and the formal fitting uncertainty, $\sigma _{\mathrm{f}}$. The total error in the flux density is then assumed to be , $\sigma _{\mathrm{t}} = (\sigma _{\mathrm{s}}^2 + \sigma _{\mathrm{f}}^2)^{0.5}$.

\section{Results}\label{sec3}
\begin{figure*}[hbt!]
\centering 
\mbox{
          \includegraphics[trim=0pt 150pt 0pt 95pt,clip,width=6.8cm,angle=90]{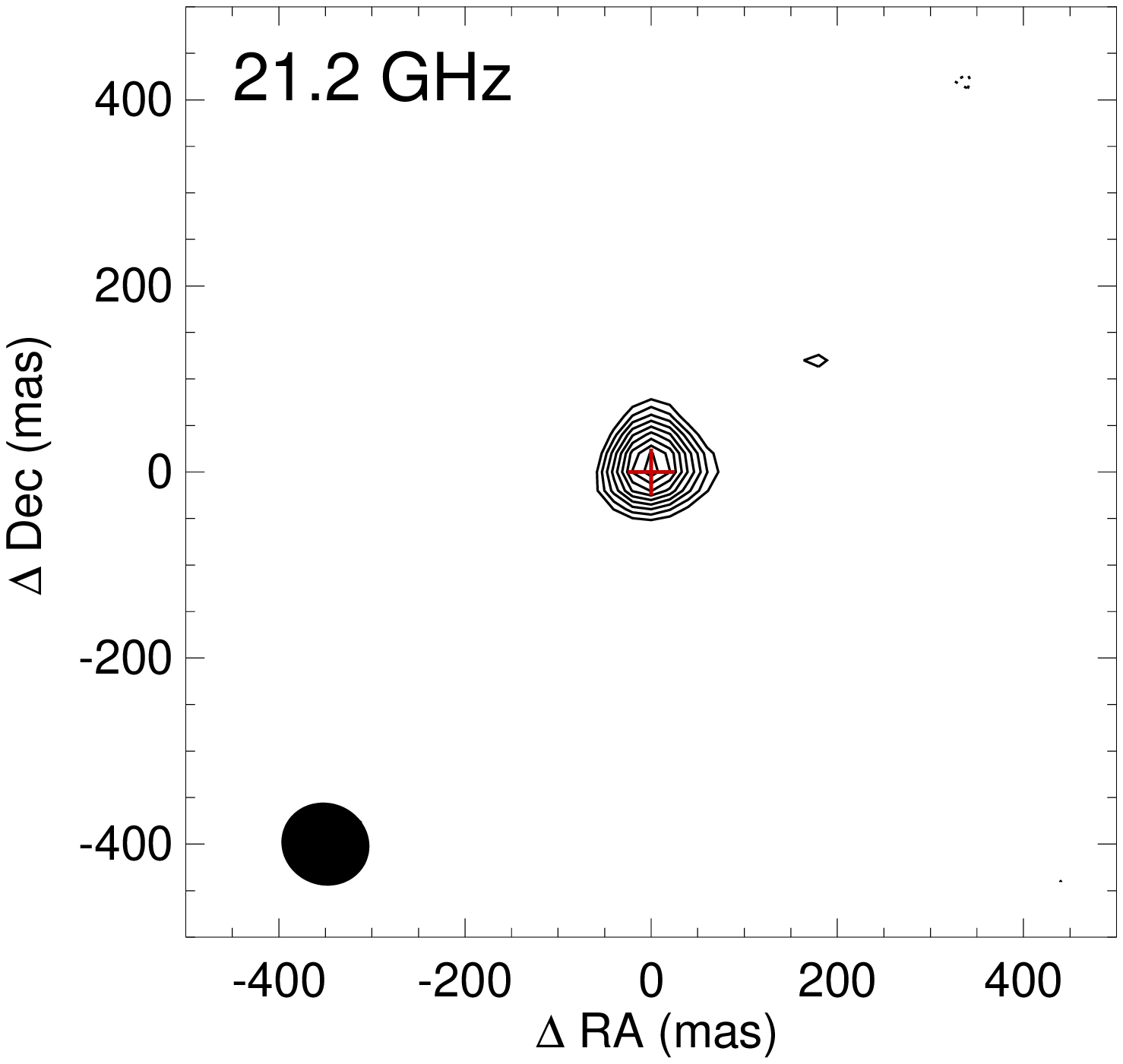}
          \includegraphics[trim=0pt 140pt 0pt 80pt,clip,width=6.8cm,angle=90]{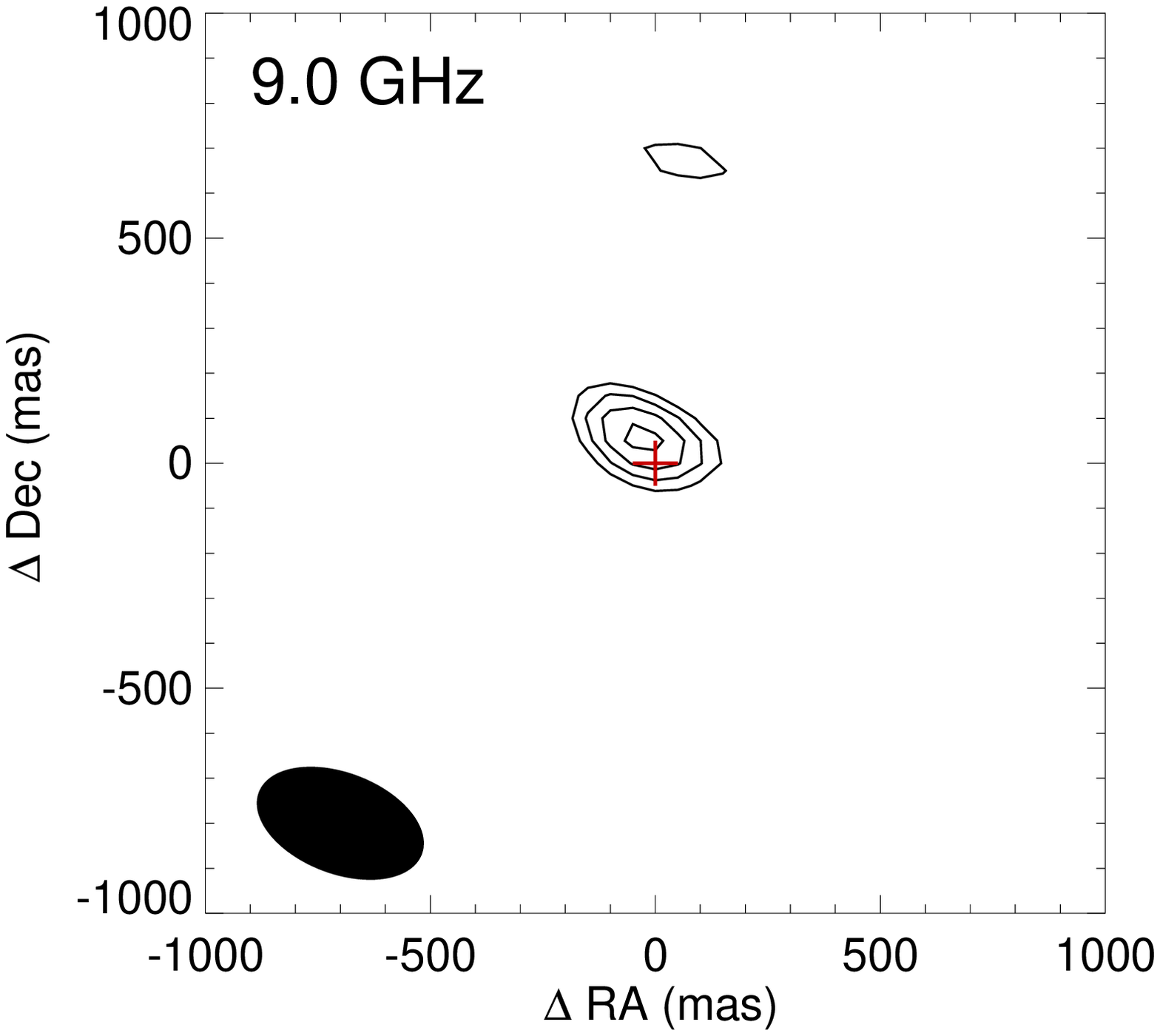}
          \includegraphics[trim=0pt 0pt 0pt 85pt,clip,width=6.8cm,angle=90]{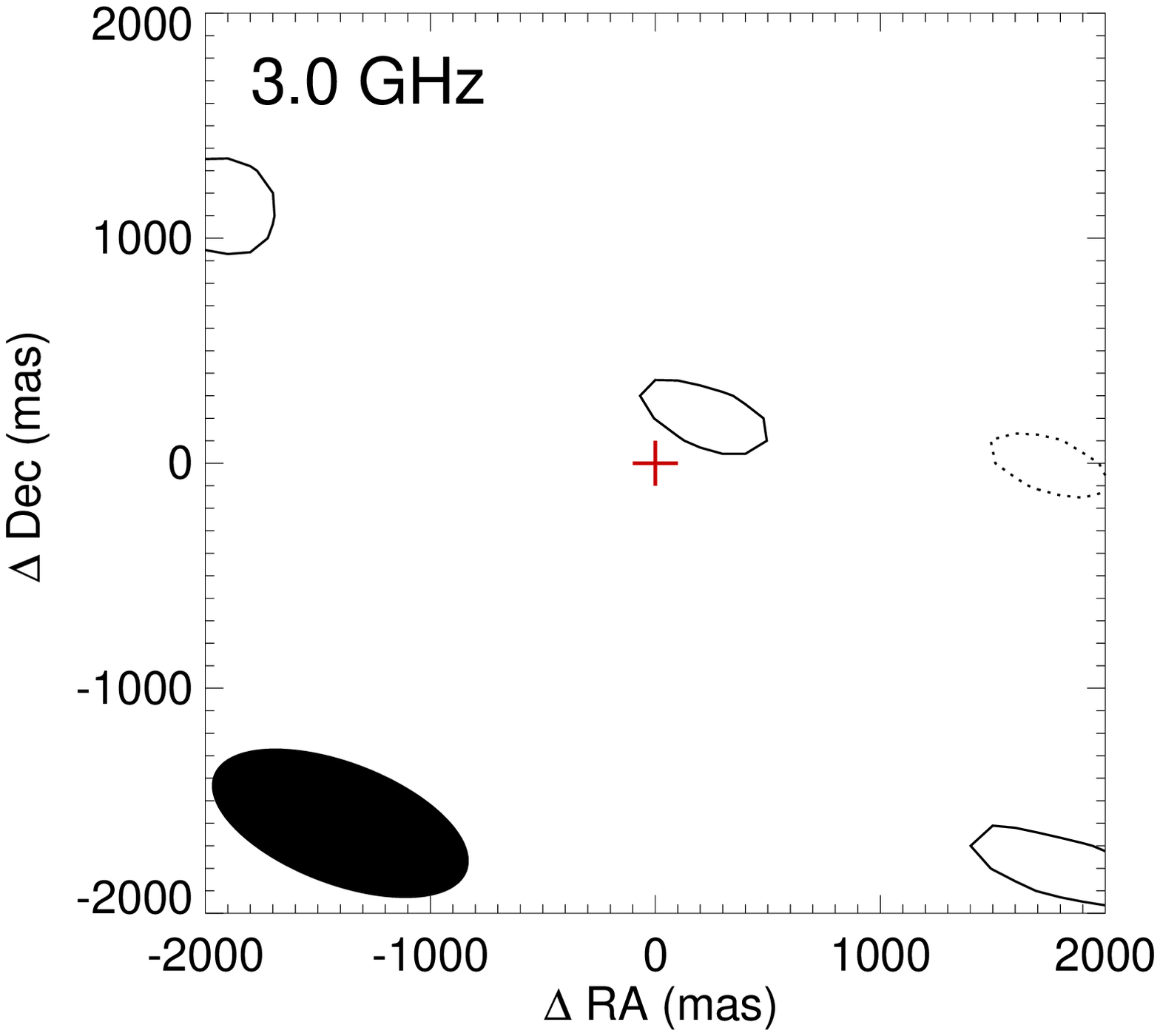}
          }
\caption{Radio images of Pollux at 21.2\,GHz (left), 9\,GHz (middle), and 3\,GHz (right). Contours are set to $(-3,3,4,...11)\times \sigma _{\mathrm{rms}}$ where $\sigma _{\mathrm{rms}}$ is the rms noise in each image. The red cross marks the expected position of the photosphere at the epoch of each observation. The restoring beam is shown in the bottom left of each image. Note the scales on the axes in each panel double in size from left to right.}
\label{fig0}
\end{figure*}

Pollux was detected (i.e., at $>3\sigma _{\mathrm{rms}}$ significance) at 21.2 and 9\,GHz in Stokes \textit{I} (total intensity) in both the point source fits to the calibrated visibilities and in the corresponding radio images, which are shown in Figure \ref{fig0}. The detections had signal-to-noise (S/N) ratios of 7 at 21.2\,GHz and 5 at 9.0\,GHz. The star was not detected at 3\,GHz or in any other Stokes parameter. The derived flux densities along with the 3\,GHz $3\sigma _{\mathrm{rms}}$ upper limit are listed in Table \ref{tab2} and are consistent with the previous non-detection of Pollux by \cite{drake_1986} who reported a $3\sigma _{\mathrm{rms}}$ upper limit of $340\,\mu$Jy at 5\,GHz. We fitted an elliptical Gaussian to the source using CASA's \textit{imfit} task to confirm that the star was unresolved at both 21.2 and 9.0\,GHz.

The expected positions of Pollux at the two observational epochs were calculated using the \textit{Hipparcos} coordinates, proper motion, and parallax \citep{van_leeuwen_2007}. The positional offset of the source could then be determined from the difference between the expected position and the actual measured position. At both 21.2 and 9\,GHz, we find that the positional offset is smaller than the noise-based uncertainty, which we define as the synthesized beam HPBW divided by $2\,\times\,\mathrm{S/N}$ of the source \citep{condon_1998}. This good agreement confirms that we have detected Pollux at 21.2 and 9\,GHz. In Figure \ref{fig0} it can be seen that a $3\sigma _{\mathrm{rms}}$ peak lies close to the expected position of Pollux at 3\,GHz. However, considering that this peak lies at a distance of almost two times the noise-based uncertainty away from the expected position, along with the fact that there are other peaks of comparable statistical significance close by, we do not attempt to consider this peak as a detection.

\begin{table*}
\caption{VLA observations of Pollux.}
\label{tab2}
\centering
\begin{tabular}{c c c c c c c c c c}
\hline\hline
Date						& Band			    & Center            & Bandwidth      		& Time on   & Restoring  	& Flux                & Brightness  \\
						    & 			        & Frequency         &      		        & Source    &  Beam HPBW       & Density                & Temperature   \\
                			&                   &  (GHz)   	     & (GHz)               	& (min)   	 & ($\arcsec \, \times \, \arcsec$)   	& ($\mu$Jy)   	    & (K)  \\
\hline
\rule{-2.6pt}{2.5ex} 2012 Oct 14  & K & 21.2 & 2.0 & 32 & 0.092\,$\times$\,0.085 & $150\pm 21$ &  $9300\pm1300$\\
                     2013 Jan 01  & X & 9.0 & 2.0 & 17 & 0.381\,$\times$\,0.218 & $43\pm 8$ &  $14800\pm2800$\\
					  2013 Jan 01  & S & 3.0 & 2.0 & 17 & 1.194\,$\times$\,0.534 & <23  &  $<71000$\\
\hline
\end{tabular}
\tablefoot{The brightness temperatures are calculated from Equation \ref{eq0}, where we have assumed that the stellar diameter is equal to the optical diameter and that the star is a uniform disk.}
\end{table*}

\section{Discussion}\label{sec4}
\subsection{Spectral indices and brightness temperatures}
Figure \ref{fig1} shows the radio spectral energy distribution for Pollux. We find that the spectral index $\alpha$ (where $S_{\nu} \propto \nu ^{\alpha}$) has a value of $\alpha = 1.5\pm 0.4$ between 9 and 21.2\,GHz, and has a value of $\alpha > 0.6\pm 0.5$ between 3 and 9\,GHz. These spectral index values are in broad agreement with the spectral indices of the disk-averaged quiet Sun over the same frequency ranges, i.e.,  $\alpha _{\odot}= 1.75$ between 6 and 400\,GHz, and $\alpha _{\odot} = 0.56$ between 0.35 and 6\,GHz \citep{benz_2009}. It can be seen that the flux densities at 9 and 21.2\,GHz are significantly above the values expected from a blackbody with a photospheric angular diameter and a temperature equal to the effective temperature. The derived brightness temperatures are listed in Table \ref{tab2} and are consistent with the solar minimum values.

\subsection{Radio emission mechanisms}
In the following sections we discuss the possible radio emission mechanisms responsible for the observed radio flux from Pollux. We note that our discussion naturally follows the discussions outlines already presented in \cite{drake_1993} and \cite{villadsen_2014}. In \cite{drake_1993}, the possible radio emission mechanisms responsible for their 8.3\,GHz detections of the subgiant star Procyon are discussed. A similar layout is applied in \cite{villadsen_2014} to discuss their 34.5\,GHz detections of three solar-type MS stars (i.e., $\tau$ Cet, $\eta$ Cas A, and 40 Eri A).
\subsubsection{Free-free emission from a chromosphere and transition region}
\begin{figure}[]
\includegraphics[trim = 0mm -100mm 0mm 0mm, clip,scale=0.39,angle=90]{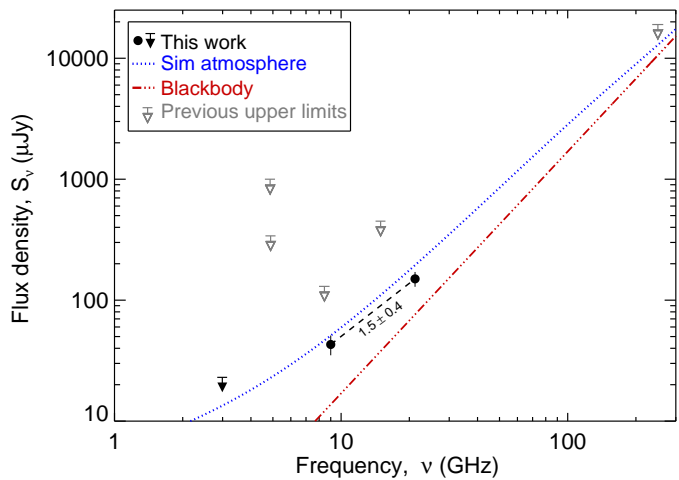}
\caption{Radio spectral energy distribution for Pollux between 1 and 300\,GHz. The red dash-dotted line is the expected blackbody emission assuming a uniform intensity disk with angular diameter $\phi =\phi _{\star}$ and brightness temperature $T_b = T_{\mathrm{eff}}$ . The black filled circles are the VLA detections at 21.2 and 9\,GHz. The blue dotted line is the expected spectral energy distribution of Pollux using the model atmosphere from \cite{sim_2001}. Our 3\,GHz upper limit is also shown along with previous upper limits taken from \cite{wendker_1995}. The black dashed line represents a least-square fit to our detections assuming that $S_{\nu} \propto \nu ^{\alpha}$.}
\label{fig1}
\end{figure}

If Pollux were a perfect blackbody with angular diameter $\phi = \phi _{\star}$ and brightness temperature $T_\mathrm{b} = T_{\mathrm{eff}}$ at any radio frequency $\nu$, then the observed flux density would be
\begin{equation}\label{eq0}
S_{\nu} = 0.18\mu \mathrm{Jy}\left(\frac{T_\mathrm{b}}{4904\,\mathrm{K}} \right)\left( \frac{\nu}{\mathrm{GHz}}\right)^2\left( \frac{\phi}{7.95\,\mathrm{mas}}\right)^2.
\end{equation}
This blackbody radio spectrum is plotted in Figure \ref{fig1} along with our flux density measurements which are clearly all in excess of this spectrum. This is because Pollux, like the Sun, has a chromosphere and a transition region. At radio frequencies, electrons and ions from these regions are a source of free-free opacity which increases as frequency decreases. This means that at lower radio frequencies an optical depth of unity is reached higher in the atmosphere where the gas/electron temperature $T_\mathrm{e} > T_{\mathrm{eff}}$ and $\phi > \phi _{\star}$, and so the flux density will be greater than that expected from a blackbody alone\footnote{At very high frequencies (i.e., $\nu \gtrsim 1\,$THz) the temperature minimum will be probed, in which case $T_\mathrm{b} < T_{\mathrm{eff}}$ and the flux density will be \textit{less} than that expected from a blackbody \citep[e.g.,][]{liseau_2013}.} \citep[e.g.,][]{cassinelli_1977}. In Figure \ref{fig1} we also include the predicted radio spectral energy distribution from the atmospheric model of Pollux by \cite{sim_2001}. This model contains a turbulently extended chromosphere, a transition region and corona and is in reasonably agreement with our measurements. We note that the semi-empirical model is not a close match to the observed \ion{H}{I} Lyman profiles, and therefore the model itself has intrinsic uncertainties. The radio emission from this model atmosphere is expected to be in excess of the blackbody emission and this excess is expected to be greater at lower frequencies, in agreement with our measurements.

\begin{figure}[]
\includegraphics[trim = 0mm 0mm 0mm 10mm, clip,scale=0.4,angle=90]{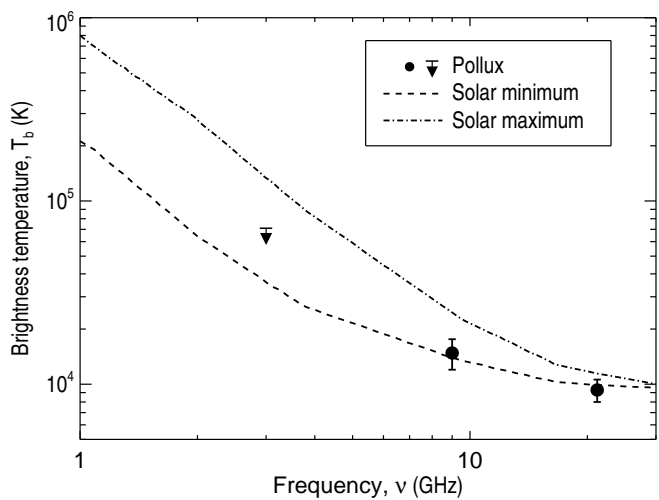}
\caption{The derived disk averaged brightness temperature of Pollux at 21.2 and 9\,GHz, along with the upper limit value at 3\,GHz. The dashed line is the solar minimum disk averaged brightness temperature while the dash dotted line is the solar maximum disk averaged brightness temperature \citep{white_2004}.}
\label{fig2}
\end{figure}

In Figure \ref{fig2} we compare the disk-averaged brightness temperatures of Pollux against the solar minimum and maximum values taken from \cite{white_2004}. The brightness temperature of the Sun changes between solar minimum and maximum mainly because of gyroresonance emission above active regions which will be discussed further in Section \ref{gyro}. At solar minimum, radio emission between 3 and 21\,GHz is dominated by optically thick thermal emission from from the upper chromosphere and lower transition region. The relationship between the radio optical depths in the quiet Sun and Pollux can be estimated from the chromospheric scaling relations of \citet{ayres_1979}. The chromospheric free-free optical depth scales as $\tau \propto n_e^2 H$, where $n_e$ is the electron density and $H$ is the density scale-height. $H$ scales as $\propto T_{\mathrm{eff}}/g_\ast$, where $g_\ast$ is the surface gravity, while $\langle n_e\rangle \propto \sqrt{g_\ast} T_{\mathrm{eff}}^{5/2}$, so that the surface gravity terms cancel \citep{harper_2013}. The remaining $T_{\mathrm{eff}}^6$ term gives $\tau _{\mathrm{Pollux}}/\tau _{\odot} \sim 0.4$ which suggests that Pollux will have a slightly lower opacity chromosphere, and for a given frequency, deeper and cooler plasma is probed. Given the similarities between the solar brightness temperatures and the values we find for Pollux, it seems likely that Pollux's radio emission also originates mainly from an upper chromospheric and lower transition region at our observing frequencies.

\subsubsection{Free-free thermal emission from a bound corona}\label{sec4.2.2}
We now examine the possibility that some of the radio emission results from free-free thermal emission from a bound (i.e., static) optically thin corona. We can immediately rule out an optically thick corona on the basis that the expected brightness temperatures would be much larger than those we find. Since the solar corona is optically thin above a few GHz, the rising spectrum (or positive spectral indices) we deduce for Pollux immediately suggests emission at 9 and 21.2\,GHz is not dominated by optically thin coronal emission, because in this case the spectrum would be almost flat (i.e., $S_{\nu} \propto \nu ^{-0.1}$).

Pollux is a weak soft X-ray source \citep{maggio_1990} and its X-ray ($5-100\,\AA$) luminosity of $10^{27.13}$\,erg\,s$^{-1}$ \citep{sanz_forcada_2011} is similar to the solar minimum value defined in \cite{judge_2003}. Thus, Pollux - with a surface area 78 times that of the Sun - has an X-ray surface flux that is about 1\% of the solar value. This soft X-ray flux originates as thermal emission from an optically thin corona and is the sum of many different thermal processes such as free-free, bound-free, and bound-bound. Knowledge of the coronal emission measure ($EM$, measured in cm$^{-3}$) allows the expected radio luminosity to be calculated using
\begin{equation}
L_{\nu} = \epsilon _{\nu}(T_e) EM
\end{equation}
where $\epsilon _{\nu}(T_e)$ is the radiative loss rate of the plasma (in erg s$^{-1}$ cm$^{3}$ Hz$^{-1}$) and is assumed to be entirely from free-free interactions. This is calculated using Equations 3.54 and 3.55 from \cite{spitzer_1978} and assuming helium to be fully ionized. Using the larger $EM$ value and the corresponding coronal temperature of $T_e = 10^{6.23}\,$K from the two temperature $EM$ fit given in \cite{sanz_forcada_2011}, we find the expected flux density contribution from Pollux's optically thin corona to be $S_{\nu} \sim 0.5\,\mu$Jy at 3\,GHz and even smaller at higher frequencies. We conclude that the free-free thermal emission from a bound corona has only a very small contribution to our radio detections.

\subsubsection{Gyroresonance emission from active regions}\label{gyro}
The non-flaring Sun displays a clear cycle in its radio flux density. This is due to coronal material above active regions becoming optically thick from the presence of hot dense plasma and strong magnetic fields and results in bright spots of coronal temperatures ($\sim10^6$\,K) appearing on a cooler chromospheric temperature ($\sim10^4$\,K) disk. As the number of active regions changes over a solar cycle, so too does the radio flux density contribution from these bright spots. The dominant source of opacity in these bright spots is free-free thermal electrons in the corona at low frequencies (i.e., $< 2\,$GHz) and gyroresonance absorption by thermal electrons in the corona between 3 and $\sim\,15$\,GHz \citep{white_1997, lee_1998}. The gyroresonance mechanism is the non-relativistic limit of the gyrosynchrotron mechanism and typically operates at low harmonics ($s=2-4$) of the electron gyrofrequency $\nu _{\mathrm{B}}$, such that
\begin{eqnarray}\label{eqml0}
\nu = s\nu_{\mathrm{B}} = 2.8sB\,\mathrm{MHz}
\end{eqnarray}
for a given magnetic field strength $B$ (in Gauss). 

We can use our VLA data to constrain the covering fraction of these coronal bright spots on Pollux, following the arguments outlined in \cite{drake_1993} and \cite{villadsen_2014}. We assume that at any given frequency the disk averaged brightness temperature $T_b $ can be modelled as a uniform disk having either a chromospheric or transition region temperature $T_{\mathrm{disk}} $, that is obscured by coronal bright spots of temperature $T_{\mathrm{cor}}$ having a covering fraction $f_{\nu}$ such that 
\begin{equation}\label{eq1}
T_b = (1 - f_{\nu})T_{\mathrm{disk}} + f_{\nu}T_{\mathrm{cor}}.
\end{equation}
The covering fraction will decrease as frequency increases since higher frequencies correspond to stronger magnetic field strengths. For example, coronal magnetic field strengths of 360, 1070 and 2520\,G would be required if gyroresonance emission in the third harmonic contributes to the radio flux densities at 3, 9, and 21.2\,GHz, respectively.
We again use $T_{\mathrm{cor}} = 10^{6.23}\,K$, while for $T_{\mathrm{disk}}$, we assume the disk integrated brightness temperature values of the quiet Sun \citep{white_2004}, since these measurements will have the least contribution from gyroresonance emission. We find that the brightness temperatures for Pollux at 9 and 21.2\,GHz are similar to solar minimum values which could imply little or no contribution from gyroresonance emission at these frequencies. Our upper limit for the the disk integrated brightness temperature of Pollux at 3\,GHz is well above the solar minimum value of $\sim 3.5\times 10^4$\,K. We can thus derive a maximum covering fraction of 2\% for coronal bright spots at 3\,GHz using Equation \ref{eq1}. For this calculation we have assumed that like the Sun, Pollux contains localized active regions, an assumption which is supported by the discovery of a large scale magnetic field in the photosphere of Pollux \citep{auriere_2009}. 

\subsubsection{Free-free emission from an ionized stellar wind}\label{sec_wind}
The unbound solar corona escapes to the interstellar medium through open magnetic field lines at a mean rate of $2.5 \times 10^{-14}\,M_{\odot}$\,yr$^{-1}$ \citep{athay_1976}, forming the solar wind. In general, the mass-loss rates from giant stars are many orders of magnitude greater than the solar value. However, these mass-loss rates are extremely poorly constrained for coronal giants like Pollux. \cite{drake_1986} were able to place $3\sigma _{\mathrm{rms}}$ upper limits of $\sim 5 \times 10^{-10}\,M_{\odot}$\,yr$^{-1}$ using their best constrained non-detections of coronal giants. 

We can use the multi-frequency VLA detections to estimate the flux density contribution from Pollux's wind and thereby constrain its mass loss rate. To do so we need an approximation of the stellar wind velocity, $v_{\mathrm{w}}$. Pollux does not show any wind scattered Mg\,II or Lyman alpha features in its \textit{HST}/STIS spectra while the C\,III 977\,\AA\ and O\,IV lines in \textit{FUSE} spectra presented in \cite{dupree_2005} are consistent with no wind scattering. Therefore Pollux does not have either a cool slow velocity wind (i.e., $v_{\mathrm{w}} = 10- 50\,$km\,s$^{-1}$) typical from cool giants or a warm intermediate velocity wind (i.e., $v_{\mathrm{w}} = 70- 150\,$km\,s$^{-1}$) typical from hybrid bright giants. Given the paucity of empirical constraints on $v_{\mathrm{w}}$, we follow \cite{drake_1986} and assume $v_{\mathrm{w}} = 0.75v_{\star}^{\mathrm{esc}}$, where $v_{\star}^{\mathrm{esc}}$ is the photospheric escape velocity. This relation is representative of the mean solar wind velocity and gives $v_{\mathrm{w}} = 215\,$km\,s$^{-1}$ for Pollux. 

\cite{drake_1986} showed that all the winds in their sample of coronal giants will be optically thin at 5\,GHz. Following their arguments, we expect Pollux's wind to be optically thin at our 3 observed radio frequencies. We can derive an upper limit to the mass-loss rate by assuming the emission at 9 and 21.2\,GHz is entirely from other sources and not from the wind. Then, by extrapolating the power law fit to the flux densities at these frequencies (i.e., using $S_{\nu} \propto \nu ^{1.5}$), we can obtain an \textit{effective} flux density value at 3.0\,GHz. The difference between our 3\,GHz upper limit for the flux density and this effective flux density is the upper limit to the flux density contribution from the wind and equates to $\sim 8\,\mu$Jy. We calculate the mass loss rate from a spherically symmetric, isothermal optically thin wind to be
\begin{eqnarray}\label{eqml2}
\dot{M}_{\mathrm{ion}} &=& 1.3\times 10^{-11}\,M _{\odot}\,\mathrm{yr}^{-1}\left( \frac{v_{\mathrm{w}}}{215\,\mathrm{km}\,\mathrm{s}^{-1}}\right)\left( \frac{d}{10.36\,\mathrm{pc}}\right) \dots \nonumber \\ && {}  \left( \frac{T_\mathrm{e}}{10^{6.23}\,\mathrm{K}}\right)^{0.22}\left( \frac{\nu}{3\,\mathrm{GHz}}\right)^{0.03}\left( \frac{R _{\star}}{8.85\,R _{\odot}}\right)^{0.5}\left( \frac{S_{\nu}}{\mu \mathrm{Jy}}\right)^{0.5}
\end{eqnarray}
where $\dot{M}_{\mathrm{ion}}$ is the ionized mass loss rate and $S_{\nu}$ is the maximum excess flux density at 3\,GHz. For coronal giants, the ionized mass loss rate will be equal to the total mass loss rate, since hydrogen and helium are fully ionized. In deriving Equation \ref{eqml2} we have computed the free-free Gaunt factor, $g_{\mathrm{ff}}$, assuming an effective power-law approximation to the appropriate hydrogen helium mix [A(He)=0.1] for coronal ionization conditions between $5\times 10^5 < T_{\mathrm{e}}(\mathrm{K}) < 10^7$ and VLA frequencies, i.e., $g_{\mathrm{ff}} = 24.1 \nu^{-0.06}T_{\mathrm{e}}^{0.06}$. In addition, we have replaced the constant $v_{\mathrm{w}}$ with the beta power-law form $v_{\mathrm{r}} = v_{\mathrm{w}}(1- R_{\star}/r)^{\beta}$, where $r$ is radial distance. We adopt $r=1.2R_{\star}$ as a limit to the integral calculation to be consistent with the chromospheric extensions inferred in \cite{berio_2011} and assume $\beta = 1$ \citep{carpenter_1995, carpenter_1999}.

From Equation \ref{eqml2} we can constrain the mass loss rate of Pollux to be $\dot{M}_{\mathrm{ion}} \leq 3.7\times 10^{-11}\,M _{\odot}$, which is about an order of magnitude smaller than previous constraints at radio wavelengths \citep{drake_1986}. The semi-empirical mass loss relation from \cite{schroder_2005} predicts a mass-loss rate for Pollux that is almost identical to the the value of our upper limit, while the scaling law from \cite{holzer_1983}, which scales the solar mass loss rate by $(R_{\star}/R_{\odot})^2(v_{\star}^{\mathrm{esc}}/v_{\odot}^{\mathrm{esc}})^{-3}$, predicts a value that is a factor of two smaller. The theoretical model for mass-loss rates of cool stars by \cite{cranmer_2011} predicts $2.3\times 10^{-12} \leq \dot{M}_{\mathrm{ion}} \leq 9.2\times 10^{-12}\,M _{\odot}$, assuming $v\,{\mathrm{sin\mathit{i}}} = 1.7\,$km\,s$^{-1}$ (S. R. Cranmer, 2016, private communication). Our upper limit is therefore in agreement with existing predictions for the mass-loss rate of Pollux. The mass loss presumably occurs in magnetically open coronal regions, which account for only a portion of the total coronal free-free radio emission - the other portion coming from magnetically closed coronal regions. Since we have shown in Section \ref{sec4.2.2} that the expected coronal free-free emission is about 0.5\,$\mu$Jy, then it is probable that the actual mass-loss rate is considerably less than our derived upper limit.

\subsubsection{Exoplanetary emission}
The magnetized planets in our solar system produce electron cyclotron maser emission at very low frequencies (i.e., $\nu <40\,$MHz) and it has long been suggested that this type of emission could potentially be detected from nearby magnetic exoplanets. Predictions for the strength of the exoplanet radio emission have been made whereby simple scaling laws known to operate in our solar system have been generalized for various stellar systems \citep[e.g.,][]{farrell_1999, stevens_2005}. Pollux is known to have a planetary companion with a minimum mass of 2.9\,$M_{\mathrm{Jupiter}}$ and a semi-major axis of 1.6\,AU \citep{reffert_2006, hatzes_2006}. If we optimistically assume that Pollux has a mass loss rate of $ 3.7\times 10^{-11}\,M _{\odot}\,\mathrm{yr}^{-1}$ (see Section \ref{sec_wind}), then following \cite{ignace_2010} we could expect a contribution from exoplanetary radio emission of $\sim 25\,\mu$Jy at 3\,GHz, where we have assumed Jupiter-like properties for the rotation rate and radius of the planet and the emission to be isotropically beamed. However, since this type of radio emission is produced at the electron cyclotron frequency which is defined in Equation \ref{eqml0} with $s=1$, the planetary magnetic field strength would need to be over 1000\,G in its polar region to be detectable at 3\,GHz and would need to be even larger to be observable at higher frequencies. Such large values are highly unlikely considering Jupiter has a value of only 14\,G and is the most magnetized planet in our solar system. We therefore do not expect to detect any exoplanetary emission in our data. Nevertheless, the larger mass loss rates of evolved stars in comparison to solar type stars could make them interesting candidates to search for exoplanetary emission at lower radio frequencies. For example, repeating the above  calculation from \cite{ignace_2010} at 150\,MHz (i.e., a more realistic exoplanet emission frequency) predicts an exoplanetary radio flux density of $\sim 500\,\mu$Jy, which would be detectable with the Low Frequency Array (LOFAR).

\section{Conclusions}\label{sec5}
The single coronal giant Pollux has been detected at 9 and 21.2\,GHz with the VLA and tight upper limits for its flux density have been found at 3\,GHz. The emission is thermal in nature and has disk-averaged brightness temperature values that are consistent with those of the quiet Sun. The origin of the emission is most likely from an optically thick upper chromosphere at 21.2 and 9\,GHz. There may also be a small contribution from other mechanisms such as gyroresonance emission above active regions, an optically thin static (i.e., bound) corona and an optically thin stellar wind, particularly at the lower frequencies. We constrain the total mass loss rate of Pollux to be less than $3.7\times 10^{-11}\,M _{\odot}$\,yr$^{-1}$, which is about an order of magnitude smaller than previous constraints for coronal giants.

The frequency range covered in this paper is crucial for understanding stellar atmospheres because many different emission processes, like those discussed in this paper, occur over this range. The next generation of radio observatories such as the Square Kilometre Array (SKA) will have the sensitivity to better distinguish between theses competing processes. For example, \cite{villadsen_2014} argued that a detection of circular polarization would favour gyroresonance emission at these frequencies. This could be important for more active coronal giants than Pollux which presumably have more active regions on their observable disks. High S/N detections at 3\,GHz and lower could trace emission from ionized stellar winds for coronal giants similar to Pollux. Indeed, detections of thermal continuum emission at 1.5 and 3\,GHz from non-cornal giants have already been used to constrain their partially ionized winds \citep{ogorman_2013}. Furthermore, observations of nearby coronal giants at lower frequencies than those presented here could potentially be also used to search for exoplanetary radio emission. The detections presented in this paper are the first detections of radio emission from a single (i.e., non-binary) coronal giant. Pollux has the largest angular diameter of all coronal giants, and so the quiescent stellar disk dominated emission from other nearby single coronal giants will presumably be more difficult to detect using existing radio interferometers. The increased sensitivity of the SKA will enable radio surveys of nearby coronal giants to be carried out and will help elucidate the properties of coronal giant atmospheres as these stars evolve. 

\begin{acknowledgements}
The data presented in this paper were obtained with the  Very Large Array (VLA) which is an instrument of the National Radio Astronomy Observatory (NRAO). The NRAO is a facility of the National Science Foundation operated under cooperative agreement by Associated Universities, Inc. This project was funded by ERC consolidator grant 614264. EOG also acknowledges support from the Irish Research Council. 
\end{acknowledgements}

%-------------------------------------------------------------------
\bibliographystyle{aa}
\bibliography{references}

\begin{thebibliography}{50}
\expandafter\ifx\csname natexlab\endcsname\relax\def\natexlab#1{#1}\fi

\bibitem[{{Athay}(1976)}]{athay_1976}
{Athay}, R.~G., ed. 1976, Astrophysics and Space Science Library, Vol.~53, {The
  solar chromosphere and corona: Quiet sun}

\bibitem[{{Auri{\`e}re} {et~al.}(2009){Auri{\`e}re}, {Wade},
  {Konstantinova-Antova}, {Charbonnel}, {Catala}, {Weiss}, {Roudier}, {Petit},
  {Donati}, {Alecian}, {Cabanac}, {van Eck}, {Folsom}, \&
  {Power}}]{auriere_2009}
{Auri{\`e}re}, M., {Wade}, G.~A., {Konstantinova-Antova}, R., {et~al.} 2009,
  \aap, 504, 231

\bibitem[{{Ayres}(1979)}]{ayres_1979}
{Ayres}, T.~R. 1979, \apj, 228, 509

\bibitem[{{Ayres} {et~al.}(1997){Ayres}, {Brown}, {Harper}, {Bennett},
  {Linsky}, {Carpenter}, \& {Robinson}}]{ayres_1997}
{Ayres}, T.~R., {Brown}, A., {Harper}, G.~M., {et~al.} 1997, \apj, 491, 876

\bibitem[{{Ayres} {et~al.}(2003){Ayres}, {Brown}, {Harper}, {Osten}, {Linsky},
  {Wood}, \& {Redfield}}]{ayres_2003}
{Ayres}, T.~R., {Brown}, A., {Harper}, G.~M., {et~al.} 2003, \apj, 583, 963

\bibitem[{{Ayres} {et~al.}(1991){Ayres}, {Fleming}, \& {Schmitt}}]{ayres_1991}
{Ayres}, T.~R., {Fleming}, T.~A., \& {Schmitt}, J.~H.~M.~M. 1991, \apjl, 376,
  L45

\bibitem[{{Ayres} {et~al.}(1998){Ayres}, {Simon}, {Stern}, {Drake}, {Wood}, \&
  {Brown}}]{ayres_1998}
{Ayres}, T.~R., {Simon}, T., {Stern}, R.~A., {et~al.} 1998, \apj, 496, 428

\bibitem[{{Benz}(2009)}]{benz_2009}
{Benz}, A.~O. 2009, Landolt B{\"o}rnstein

\bibitem[{{Berio} {et~al.}(2011){Berio}, {Merle}, {Th{\'e}venin}, {Bonneau},
  {Mourard}, {Chesneau}, {Delaa}, {Ligi}, {Nardetto}, {Perraut}, {Pichon},
  {Stee}, {Tallon-Bosc}, {Clausse}, {Spang}, {McAlister}, {ten Brummelaar},
  {Sturmann}, {Sturmann}, {Turner}, {Farrington}, \& {Goldfinger}}]{berio_2011}
{Berio}, P., {Merle}, T., {Th{\'e}venin}, F., {et~al.} 2011, \aap, 535, A59

\bibitem[{{Carpenter} {et~al.}(1999){Carpenter}, {Robinson}, {Harper},
  {Bennett}, {Brown}, \& {Mullan}}]{carpenter_1999}
{Carpenter}, K.~G., {Robinson}, R.~D., {Harper}, G.~M., {et~al.} 1999, \apj,
  521, 382

\bibitem[{{Carpenter} {et~al.}(1995){Carpenter}, {Robinson}, \&
  {Judge}}]{carpenter_1995}
{Carpenter}, K.~G., {Robinson}, R.~D., \& {Judge}, P.~G. 1995, \apj, 444, 424

\bibitem[{{Cassinelli} \& {Hartmann}(1977)}]{cassinelli_1977}
{Cassinelli}, J.~P. \& {Hartmann}, L. 1977, \apj, 212, 488

\bibitem[{{Condon} {et~al.}(1998){Condon}, {Cotton}, {Greisen}, {Yin},
  {Perley}, {Taylor}, \& {Broderick}}]{condon_1998}
{Condon}, J.~J., {Cotton}, W.~D., {Greisen}, E.~W., {et~al.} 1998, \aj, 115,
  1693

\bibitem[{{Cranmer} \& {Saar}(2011)}]{cranmer_2011}
{Cranmer}, S.~R. \& {Saar}, S.~H. 2011, \apj, 741, 54

\bibitem[{{Drake} \& {Linsky}(1986)}]{drake_1986}
{Drake}, S.~A. \& {Linsky}, J.~L. 1986, \aj, 91, 602

\bibitem[{{Drake} {et~al.}(1993){Drake}, {Simon}, \& {Brown}}]{drake_1993}
{Drake}, S.~A., {Simon}, T., \& {Brown}, A. 1993, \apj, 406, 247

\bibitem[{{Dupree} {et~al.}(2005){Dupree}, {Lobel}, {Young}, {Ake}, {Linsky},
  \& {Redfield}}]{dupree_2005}
{Dupree}, A.~K., {Lobel}, A., {Young}, P.~R., {et~al.} 2005, \apj, 622, 629

\bibitem[{{Farrell} {et~al.}(1999){Farrell}, {Desch}, \&
  {Zarka}}]{farrell_1999}
{Farrell}, W.~M., {Desch}, M.~D., \& {Zarka}, P. 1999, \jgr, 104, 14025

\bibitem[{{Gray}(2014)}]{gray_2014}
{Gray}, D.~F. 2014, \apj, 796, 88

\bibitem[{{Gray} {et~al.}(2003){Gray}, {Corbally}, {Garrison}, {McFadden}, \&
  {Robinson}}]{gray_2003}
{Gray}, R.~O., {Corbally}, C.~J., {Garrison}, R.~F., {McFadden}, M.~T., \&
  {Robinson}, P.~E. 2003, \aj, 126, 2048

\bibitem[{{Haisch} {et~al.}(1991){Haisch}, {Schmitt}, \& {Rosso}}]{haisch_1991}
{Haisch}, B., {Schmitt}, J.~H.~M.~M., \& {Rosso}, C. 1991, \apjl, 383, L15

\bibitem[{{Harper} {et~al.}(2013){Harper}, {O'Riain}, \& {Ayres}}]{harper_2013}
{Harper}, G.~M., {O'Riain}, N., \& {Ayres}, T.~R. 2013, \mnras, 428, 2064

\bibitem[{{Hatzes} {et~al.}(2006){Hatzes}, {Cochran}, {Endl}, {Guenther},
  {Saar}, {Walker}, {Yang}, {Hartmann}, {Esposito}, {Paulson}, \&
  {D{\"o}llinger}}]{hatzes_2006}
{Hatzes}, A.~P., {Cochran}, W.~D., {Endl}, M., {et~al.} 2006, \aap, 457, 335

\bibitem[{{Hatzes} {et~al.}(2012){Hatzes}, {Zechmeister}, {Matthews},
  {Kuschnig}, {Walker}, {D{\"o}llinger}, {Guenther}, {Moffat}, {Rucinski},
  {Sasselov}, \& {Weiss}}]{hatzes_2012}
{Hatzes}, A.~P., {Zechmeister}, M., {Matthews}, J., {et~al.} 2012, \aap, 543,
  A98

\bibitem[{{H{\"o}gbom}(1974)}]{hogbom_1974}
{H{\"o}gbom}, J.~A. 1974, \aaps, 15, 417

\bibitem[{{Holzer} {et~al.}(1983){Holzer}, {Fla}, \& {Leer}}]{holzer_1983}
{Holzer}, T.~E., {Fla}, T., \& {Leer}, E. 1983, \apj, 275, 808

\bibitem[{{Huensch} {et~al.}(1996){Huensch}, {Schmitt}, {Schroeder}, \&
  {Reimers}}]{huensch_1996}
{Huensch}, M., {Schmitt}, J.~H.~M.~M., {Schroeder}, K.-P., \& {Reimers}, D.
  1996, \aap, 310, 801

\bibitem[{{Ignace} {et~al.}(2010){Ignace}, {Giroux}, \&
  {Luttermoser}}]{ignace_2010}
{Ignace}, R., {Giroux}, M.~L., \& {Luttermoser}, D.~G. 2010, \mnras, 402, 2609

\bibitem[{{Judge} {et~al.}(2003){Judge}, {Solomon}, \& {Ayres}}]{judge_2003}
{Judge}, P.~G., {Solomon}, S.~C., \& {Ayres}, T.~R. 2003, \apj, 593, 534

\bibitem[{{Lee} {et~al.}(1998){Lee}, {McClymont}, {Miki{\'c}}, {White}, \&
  {Kundu}}]{lee_1998}
{Lee}, J., {McClymont}, A.~N., {Miki{\'c}}, Z., {White}, S.~M., \& {Kundu},
  M.~R. 1998, \apj, 501, 853

\bibitem[{{Linsky} \& {Gary}(1983)}]{linsky_1983}
{Linsky}, J.~L. \& {Gary}, D.~E. 1983, \apj, 274, 776

\bibitem[{{Liseau} {et~al.}(2013){Liseau}, {Montesinos}, {Olofsson}, {Bryden},
  {Marshall}, {Ardila}, {Bayo Aran}, {Danchi}, {del Burgo}, {Eiroa}, {Ertel},
  {Fridlund}, {Krivov}, {Pilbratt}, {Roberge}, {Th{\'e}bault}, {Wiegert}, \&
  {White}}]{liseau_2013}
{Liseau}, R., {Montesinos}, B., {Olofsson}, G., {et~al.} 2013, \aap, 549, L7

\bibitem[{{Maggio} {et~al.}(1990){Maggio}, {Vaiana}, {Haisch}, {Stern},
  {Bookbinder}, {Harnden}, \& {Rosner}}]{maggio_1990}
{Maggio}, A., {Vaiana}, G.~S., {Haisch}, B.~M., {et~al.} 1990, \apj, 348, 253

\bibitem[{{McMullin} {et~al.}(2007){McMullin}, {Waters}, {Schiebel}, {Young},
  \& {Golap}}]{mcmullin_2007}
{McMullin}, J.~P., {Waters}, B., {Schiebel}, D., {Young}, W., \& {Golap}, K.
  2007, in Astronomical Society of the Pacific Conference Series, Vol. 376,
  Astronomical Data Analysis Software and Systems XVI, ed. R.~A. {Shaw},
  F.~{Hill}, \& D.~J. {Bell}, 127

\bibitem[{{Nordgren} {et~al.}(2001){Nordgren}, {Sudol}, \&
  {Mozurkewich}}]{nordgren_2001}
{Nordgren}, T.~E., {Sudol}, J.~J., \& {Mozurkewich}, D. 2001, \aj, 122, 2707

\bibitem[{{O'Gorman} {et~al.}(2013){O'Gorman}, {Harper}, {Brown}, {Drake}, \&
  {Richards}}]{ogorman_2013}
{O'Gorman}, E., {Harper}, G.~M., {Brown}, A., {Drake}, S., \& {Richards}, A.
  M.~S. 2013, \aj, 144, 36

\bibitem[{{Parker}(1970)}]{parker_1970}
{Parker}, E.~N. 1970, \araa, 8, 1

\bibitem[{{Reffert} {et~al.}(2006){Reffert}, {Quirrenbach}, {Mitchell},
  {Albrecht}, {Hekker}, {Fischer}, {Marcy}, \& {Butler}}]{reffert_2006}
{Reffert}, S., {Quirrenbach}, A., {Mitchell}, D.~S., {et~al.} 2006, \apj, 652,
  661

\bibitem[{{Sanz-Forcada} {et~al.}(2011){Sanz-Forcada}, {Micela}, {Ribas},
  {Pollock}, {Eiroa}, {Velasco}, {Solano}, \&
  {Garc{\'{\i}}a-{\'A}lvarez}}]{sanz_forcada_2011}
{Sanz-Forcada}, J., {Micela}, G., {Ribas}, I., {et~al.} 2011, \aap, 532, A6

\bibitem[{{Schr{\"o}der} \& {Cuntz}(2005)}]{schroder_2005}
{Schr{\"o}der}, K.-P. \& {Cuntz}, M. 2005, \apjl, 630, L73

\bibitem[{{Sim}(2001)}]{sim_2001}
{Sim}, S.~A. 2001, \mnras, 326, 821

\bibitem[{{Spitzer}(1978)}]{spitzer_1978}
{Spitzer}, L. 1978, {Physical processes in the interstellar medium}

\bibitem[{{Stevens}(2005)}]{stevens_2005}
{Stevens}, I.~R. 2005, \mnras, 356, 1053

\bibitem[{{Takeda} {et~al.}(2008){Takeda}, {Sato}, \& {Murata}}]{takeda_2008}
{Takeda}, Y., {Sato}, B., \& {Murata}, D. 2008, \pasj, 60, 781

\bibitem[{{van Leeuwen}(2007)}]{van_leeuwen_2007}
{van Leeuwen}, F. 2007, \aap, 474, 653

\bibitem[{{Villadsen} {et~al.}(2014){Villadsen}, {Hallinan}, {Bourke},
  {G{\"u}del}, \& {Rupen}}]{villadsen_2014}
{Villadsen}, J., {Hallinan}, G., {Bourke}, S., {G{\"u}del}, M., \& {Rupen}, M.
  2014, \apj, 788, 112

\bibitem[{{Wendker}(1995)}]{wendker_1995}
{Wendker}, H.~J. 1995, \aaps, 109, 177

\bibitem[{{White}(2004)}]{white_2004}
{White}, S.~M. 2004, \nar, 48, 1319

\bibitem[{{White} \& {Kundu}(1997)}]{white_1997}
{White}, S.~M. \& {Kundu}, M.~R. 1997, \solphys, 174, 31

\bibitem[{{Wolff} \& {Simon}(1997)}]{wolff_1997}
{Wolff}, S. \& {Simon}, T. 1997, \pasp, 109, 759

\end{thebibliography}

\end{document}